\documentclass[iop, apjl, twocolappendix, twocolumn]{aastex631}
\usepackage{graphicx}
\usepackage{bm}
\usepackage{amsmath,amssymb}
\usepackage{booktabs}
\usepackage{color}
\usepackage{units}
\usepackage[normalem]{ulem}

\usepackage[normalem]{ulem}

\newcommand{\asharp}{A$^\sharp$~}
\newcommand{\nNSBH}{200~}

\hypersetup{colorlinks=true, citecolor=teal, urlcolor=teal, linkcolor=teal}

\begin{document}

\title{Inferring jet physics from neutron star---black hole mergers with gravitational waves}

\correspondingauthor{Teagan Clarke}
\email{teagan.clarke@monash.edu}

\author[0000-0002-6714-5429]{Teagan A. Clarke}
\affil{School of Physics and Astronomy, Monash University, VIC 3800, Australia}
\affil{OzGrav: The ARC Centre of Excellence for Gravitational-wave Discovery, Clayton, VIC 3800, Australia}

\author[0000-0003-3763-1386]{Paul D. Lasky}
\affil{School of Physics and Astronomy, Monash University, VIC 3800, Australia}
\affil{OzGrav: The ARC Centre of Excellence for Gravitational-wave Discovery, Clayton, VIC 3800, Australia}

\author[0000-0002-4418-3895]{Eric Thrane}
\affil{School of Physics and Astronomy, Monash University, VIC 3800, Australia}
\affil{OzGrav: The ARC Centre of Excellence for Gravitational-wave Discovery, Clayton, VIC 3800, Australia}

\date{\today}

\begin{abstract}
Neutron star---black hole (NSBH) mergers that undergo tidal disruption may launch jets that could power a gamma-ray burst.  
We use a population of simulated NSBH systems to measure jet parameters from the gravitational waves emitted by these systems. 
The conditions during the tidal disruption and merger phase required to power a gamma-ray burst are uncertain. 
It is likely that the system must achieve some minimum remnant baryonic mass after the merger before a jet can be launched to power a gamma-ray burst. 
Assuming a fiducial neutron star equation of state, we show how Bayesian hierarchical inference can be used to infer the minimum remnant mass required to launch a gamma-ray burst jet as well as features of the opening angle distribution of the gamma-ray burst jets. 
We find that with \nNSBH NSBH observations, we can measure the minimum disk mass to within $\unit[0.01]{M_\odot}$ at 90\% credibility. 
We simultaneously infer the  gamma-ray burst opening angle to within $\unit[13]{^\circ}$ at 90\% credibility. 
We conclude that upcoming upgrades to the LIGO observatories may provide important new insights into the physics of NSBH jets.
\end{abstract}

\section{Introduction}
\label{sec:intro}

\begin{figure*}
    \centering
    \includegraphics[trim={0.1cm 0.4cm 0.2cm 0.2cm},clip, width=\textwidth]{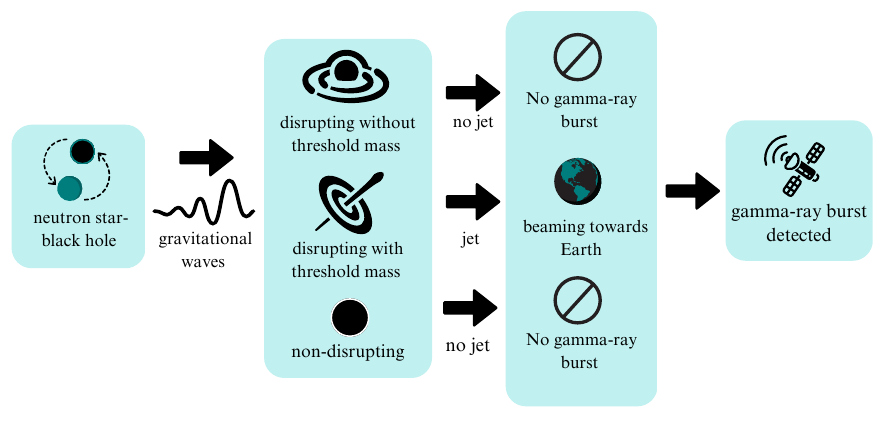}
    \caption{Schematic explaining the possible fates of an NSBH merger, depending on the binary parameters. We are interested in answering the question: what threshold remnant baryonic mass is required to launch a GRB jet? We consider the remnant mass and orientation of the system as the only two requirements to measure a GRB from an NSBH merger, however we note that in reality there may be other requirements that need to be satisfied to successfully launch a GRB jet.}
    \label{fig:flowchart}
\end{figure*}

Neutron star---black hole (NSBH) mergers probably make up at least three of the $\gtrsim$ 90 gravitational-wave events published by the LIGO--Virgo--KAGRA (LVK) collaboration \citep{GWTC2, O3b_population, gwtc3, nsbh_discovery_2021, Gw230529_2024}.\footnote{We refer to the two NSBH detections described in \cite{nsbh_discovery_2021} and the recent NSBH binary merger GW230529 \citep{Gw230529_2024}, although 1--2 more NSBH mergers could be counted in the third gravitational-wave catalog \citep{gwtc3}.} 

If the black hole in an NSBH system is sufficiently light or spinning rapidly, the neutron star may tidally disrupt. The disruption could provide a progenitor for a gamma-ray burst (GRB) \citep[e.g.,][]{Mochkovitch_1993,Janka_1999,NAKAR_2007} by launching a relativistic jet. To date, the only confirmed GRB+gravitational-wave multi-messenger detection was the binary neutron star merger GW170817 \citep{BNS_discovery_170817, Abbott2017_multi} and the corresponding GRB 170817A \citep{Goldstein2017}.

Tidal disruption alone is probably not sufficient for the system to launch a jet and for the jet to be detected on Earth as a GRB. 
We consider two requirements for an NSBH merger to launch a jet that can be measured by gamma-ray observatories at Earth. First, the NSBH must have disrupted and produced a remnant mass large enough to launch a jet that is powered for long enough to break through the merger ejecta material. The required remnant mass is likely on the order of a few percent of a solar mass, but this threshold mass is not precisely known \citep[e.g.,][]{Lee2007}. Typical disk mass threshold values used in the literature lie between 0.03 and 0.08 solar masses \citep{Pannarale2014, Stone2013,  Zappa_2019}. 
Second, the GRB jet must be launched with an opening angle and inclination that makes it possible to be viewed from Earth.
It is likely that the binary mergers' inclination angle must be close to on-axis, i.e, the orbital angular momentum axis is pointed towards Earth, for the GRB jet to be observed \citep[e.g.,][]{Chen2013, O'Connor2024}. However, the discovery of a GRB with the off-axis multi-messenger signal GW170817 \citep{BNS_discovery_170817, Abbott2017_multi, Finstad2018, Ghirlanda2019} means that jets can be observed off-axis to some degree, perhaps at a reduced luminosity and delayed rise time compared to an on-axis counterpart \citep[e.g.,][]{Granot2002}.
The jet opening angle distribution is also unknown as it depends on uncertain jet physics. 
We summarize the possible outcomes for NSBH mergers in Figure~\ref{fig:flowchart}, highlighting the requirements that need to be satisfied before a GRB can be observed. The strict physical requirements for disruption means that disrupting binaries may be a small minority of NSBH systems \citep[e.g.,][]{Fragione_2021, Zhu2022, Drozda2022, Biscoveanu_2023}, however, future detectors such as Cosmic Explorer \citep{CE} and Einstein Telescope \citep{ET} will allow us to observe a population of potentially disrupting binaries \citep[e.g.,][]{Gupta2023}, some with multi-messenger counterparts. 

In this \textit{Paper}, we describe how data from upcoming gravitational-wave observatories can be used to infer the minimum remnant mass and jet opening angle required to detect a GRB jet with minimal assumptions about the jet physics.  
We examine a possible population of NSBH mergers that could be detected by a proposed upgrade to the LIGO detectors called \asharp \citep{ligo_A_sharp}. 
\asharp will improve on the LVK design sensitivity \citep{adv_ligo_2015, AdvancedVirgo, 2020_Kagra} by a factor of approximately two in strain and is intended as a precursor instrument to Cosmic Explorer \citep{CE}, and is slated to observe in the early 2030s. 

We show that we can resolve the minimum remnant mass to launch jets and constrain the jet opening angle with a population of NSBH mergers observed by \asharp in approximately one year, provided some mergers have a multi-messenger counterpart. 
We show that our framework is feasible for populations where we assume all gamma-ray burst jets have the same opening angle and if we assume they are drawn from a distribution of opening angles.
We note that in some systems a jet could be launched without an accompanying GRB, but for this work we exclusively consider GRB jets.

The remainder of this \textit{Paper} is structured as follows. 
In Section~\ref{sec:population} we describe our simulated population that we use to infer the threshold remnant mass. In Section~\ref{sec:maths} we describe the mathematical framework for our hierarchical inference model. 
We describe our results in Section~\ref{sec:results} and discuss the implications for these results in Section~\ref{sec:discuss}. 

\section{fiducial population}
\label{sec:population}
We simulate a population of NSBH mergers injected into the LIGO Hanford and Livingston observatories at \asharp sensitivity. The parameters of our NSBH population are based on the population inference results from GW230529 \citep{Gw230529_2024, GW230529_data}, inferred using the aligned-spin \texttt{NSBH-pop} model \citep{Biscoveanu_2023}. We list the population parameters we use in Table~\ref{tab:pop_params}.
We inject \nNSBH events at \asharp sensitivity, which is consistent with the detections \asharp will make in one year of observing with network signal-to-noise ratio $> 10$ \citep{Gupta2023, Gupta2023b}, out to a luminosity distance of $\unit[1000]{Mpc}$ \citep[e.g.,][]{Chen2013}. We assume a local merger rate of $\unit[45]{Gpc^{-3}yr^{-1}}$ , which is consistent with the NSBH merger rate inferred for GWTC-3 and the GW230529 population inference results \citep{nsbh_discovery_2021, O3b_population, Biscoveanu_2023, Gw230529_2024, GW230529_data}. We generate events uniformly in co-moving volume. 

We inject and recover the signals using the Bayesian inference library \texttt{Bilby} \citep{Ashton_2019_bilby, Romero_Shaw_bilby} and the \texttt{dynesty} nested sampler \citep{dynesty}. 
We inject the signals into Gaussian noise colored by the \asharp amplitude spectral density for two observatories located at the sites of LIGO Hanford and Livingston.\footnote{We use the \texttt{Asharp\_strain.txt} amplitude spectral density curve taken from https://dcc.ligo.org/LIGO-T2300041/public \citep{ALIGO_curves}.}
We sample from uniform priors in the chirp mass, mass ratio, aligned spin components and luminosity distance.\footnote{We sample with 1000 live points, phase and time marginalization turned on, and a stopping criterion of $ \Delta \text{log}\mathcal{Z} < 0.1$, where $\mathcal{Z}$ is the Bayesian evidence.}
We use the binary black hole waveform approximant \texttt{IMRPhenomPv2} \citep{Hannam2014} and speed up our inference by employing reduced-order-quadrature in our likelihood evaluations \citep{Canizares2015, Smith2016, Morisaki2023}. 

We calculate the remnant mass posteriors of the population in post-processing. We use the fitting formulae, including the spin dependent properties of neutron stars \citep{Cipolletta2015, Breu2016, Foucart2018}, which were used in \cite[e.g.,][]{Biscoveanu_2023, Gw230529_2024} to calculate the remnant baryonic mass remaining outside of the black hole following the merger. Since the remnant mass equation requires the neutron star compactness, we use \texttt{SLY9} \citep{Douchin2001, Danielewicz2009, Gulminelli2015} as a representative equation of state that is consistent with astronomical measurements of neutron stars \cite[e.g,][]{Legred_2021}.

Finally, we introduce an electromagnetic measurement to infer if a jet is observed. We choose the minimum remnant mass to launch a jet, $M_\text{min} = \unit[0.03]{M_\odot}$ and jet opening angles, $\Theta_\text{j} =\unit[35]{^\circ}$. 
These parameters are chosen optimistically but are within the current uncertainties for jet-launching physics; e.g., \citet{Pannarale2014} use a threshold disk mass of $\unit[0.03]{M_\odot}$, while $\unit[35]{^\circ}$ is consistent with estimates of the upper limit of detecting emission from short GRBs \citep[e.g.,][]{Finstad2018, Mazwi2024}. We also note that jets from NSBH mergers may be less collimated than those from binary neutron stars, which could allow NSBH jets to be observed from larger viewing angles \citep[e.g.,][]{Sarin2022}.
We first test our framework assuming that all GRB jets have the same opening angle and later relax this assumption to include a distribution of jet opening angles in our model.

\begin{figure}
    \centering
    \includegraphics[trim={0.4cm 0.35cm 0.cm 0.1cm},clip, width=\columnwidth]{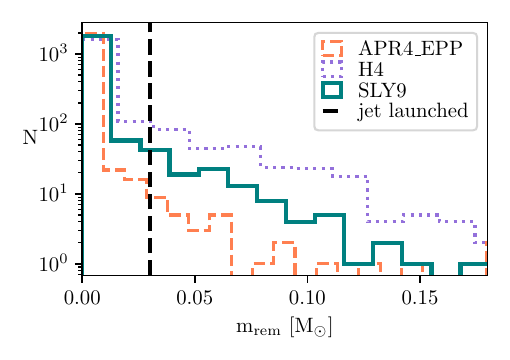}
    \caption{Remnant masses for our population of \nNSBH NSBH systems. We show the distribution we use as the ``true'' distribution, using the \texttt{SLY9} equation of state \citep{Douchin2001, Danielewicz2009, Gulminelli2015} in solid teal. We show the distributions one would get if assuming a softer equation of state (\texttt{APR4\_EPP} \citep{Akmal1998} shown by the orange dashed curve) or stiffer equation of state (\texttt{H4} \citep{Lackey2006} shown by the purple dotted curve). The equation of state has a relatively minor effect on the distribution of remnant masses, although the stiffest equation of state increases the proportion of disrupting systems with non-zero remnant masses. We mark our chosen minimum remnant mass to power a GRB ($M_\text{min}$) with a dashed black line.}
    \label{fig:remnants}
\end{figure}

We assume a top-hat jet structure where all systems that satisfy our population constraints are detected as GRBs. We only consider whether a GRB was detected using our model with a yes-no boolean framework and do not employ any modeling on the GRB energy or structure. 
Figure~\ref{fig:remnants} shows the distribution of remnant masses we obtain with our chosen population, using the \texttt{SLY9} equation of state, compared against stiffer and softer equations of state. In our model, 38 ($\approx 19\%$) of the NSBH mergers are disrupting with remnant masses $> \unit[0]{M_\odot}$ and 9 ($\approx 5 \%$) of them clear our threshold disk mass of $\unit[0.03]{M_\odot}$ to launch a GRB jet. Accounting for the maximum viewing angle of $\unit[35]{^\circ}$ yields 5 ($\approx 3 \%$) systems that launch a jet detectable at Earth.
We discuss the implications of our model assumptions in Section~\ref{sec:discuss}. 

\begin{table*}
    \centering
    \begin{tabular}{ccc}
       parameter  & description  & value \\
         \hline
         $\alpha$ & black hole mass power-law index & 1.8 \\
         $m_\text{BH,min}$ & minimum black hole mass  & $\unit[3.4]{M_\odot}$\\
        $m_\text{BH,max}$ & maximum black hole mass  & $\unit[9.5]{M_\odot}$ \\ 
         $m_\text{NS,max}$ & maximum neutron star mass & $\unit[2.0]{M_\odot}$ \\
         $\mu$             & mass ratio mean       & 0.32 \\
         $\sigma$          & mass ratio standard deviation & 0.54 \\
         $\alpha_\chi$      & beta distribution $\alpha$ shape parameter for black hole spin & 1.4 \\
         $\beta_\chi$ & beta distribution $\beta$ shape parameter for black hole spin & 7.0
         
    \end{tabular}
    \caption{population parameters for our injected population of NSBH binaries. We select population parameters from the \texttt{NSBH\_pop} model posterior distribution sourced from the GW230529 data release samples \citep{Gw230529_2024, GW230529_data}.}
    \label{tab:pop_params}
\end{table*}

\section{hierarchical model}
\label{sec:maths}
Using the NSBH population model, we calculate a posterior distribution for the minimum remnant mass required to launch a jet. 
We employ a hierarchical inference framework for this problem.
Our simulated data consists of gravitational-wave posteriors, and yes-no observations from our hypothetical gamma-ray observatory. 
We define hyper-parameters $\Lambda$, which include the threshold remnant mass $M_\text{min}$ and the distribution of jet opening angles $\Theta_j$. 
The posterior probability of a sample in our hyper-parameter-space $\Lambda$ given the gravitational-wave data $d$ and electromagnetic observation data $j$, is: 
\begin{widetext}
\begin{equation}
    p(\Lambda|d,j, W)= \pi(\Lambda) \prod_i^N  \int d\theta_i \int d\Theta^i_j \ \mathcal{L}(d_i|\theta_i) \\
    \mathcal{L}(j_i|\theta_i, \Theta_j, \Lambda) 
    \pi(\theta_i | W) \pi(\Theta_j).
\label{eq:p_lambda}
\end{equation}
Here $d$ is the gravitational-wave data and  $j$ is the GRB data, which can take values of yes or no. 
$\Theta_j$ is the jet opening angle distribution, and $\pi(\Theta_j)$ is the prior probability on this distribution. We firstly assume $\Theta_j$ as a delta function where all jets have the same opening angle. We later test a distribution of jet opening angles where we draw opening angles from a truncated Gaussian distribution between $\unit[2]{^\circ}$ and $\unit[50]{^\circ}$ with mean $\unit[25]{^\circ}$ and width $\unit[5]{^\circ}$ \citep{Biscoveanu2020}.
Meanwhile, $\theta_i$ refers to the NSBH binary parameters $\theta$ for event $i$, and $W$ is the population model that the individual events are drawn from. The prior probability of $\theta_i$ given the population model $W$ is given by $\pi(\theta_i | W)$. 
The likelihood $\mathcal{L}(d_i|\theta_i)$ is the likelihood obtained from gravitational-wave parameter estimation, for which we employ the standard Whittle likelihood function \citep[e.g.,][]{Thrane2019}. Finally, we construct the likelihood of detecting a GRB jet, $\mathcal{L}(j|\theta_i^k, \Lambda)$ as a boolean, similar to the method of \cite{Mancarella2024}, characterized by Table~\ref{tab:jet_likeihood}, representing a top-hat jet model where a GRB jet is either measured or not measured, depending on the threshold jet mass and inclination angle of the system. We note that our measurements for the system viewing angles only come from the gravitational-wave measurements. One could constrain the viewing angle with electromagnetic data as well and combine the posteriors, although care must be taken to avoid biasing the measurement \citep[e.g.,][]{Muller2024}. 
The likelihood of the gravitational-wave data can be written as 
\begin{equation}
    {\cal L}(d_i|\theta_i) =
    p(\theta_i|d_i) \frac{{\cal Z}_\varnothing}{\pi(\theta_i|\varnothing)}.
\end{equation}
We note that the posterior samples $p(\theta_i | d_i)$ are drawn from some default prior $\pi(\theta_i|\varnothing)$. Using this default prior we obtain the evidence $\cal{Z}_\varnothing$. 
Plugging this into Eq.~\ref{eq:p_lambda} we obtain
\begin{equation}
    p(\Lambda|d,j, W) \propto  \pi(\Lambda) \prod_i^N  \int d\theta_i \int d\Theta^i_j \, p(\theta_i|d_i) \frac{{\cal Z}_\varnothing}{\pi(\theta_i|\varnothing)} \\ 
    \mathcal{L}(j_i|\theta_i, \Theta_j,  \Lambda)\pi(\theta_i | W)\pi(\Theta_j).
\label{eq:p_lambda}
\end{equation}

Using that Eq.~\ref{eq:p_lambda} now takes the form
\begin{align}
    \int dx \, p(x) f(x) = \sum_k f(x_k),
\end{align}
where $p(x)$, is a posterior distribution ($p(\theta_i|d_i)$), $x_k$ are the samples drawn from $p(x_k)$ and $f(x)$ is everything else in the integrand, we obtain the following posterior probability distribution for $\Lambda$: 
\begin{equation}
    p(\Lambda|d, j, W) = \pi(\Lambda)\prod^N_i \frac{\mathcal{Z}_i}{n_i} \, \pi(\Theta^i_j) \sum^{n_i}_k \mathcal{L}(j|\theta_i^k, \Theta_j, \Lambda) \frac{\pi(\theta_i^k|W)}{\pi(\theta_i^k | \varnothing)}.
\label{eq:final_p_lambda}
\end{equation}

\end{widetext}

\begin{table}
    \centering
    \begin{tabular}{c | c c}
& $i_v \leq \Theta_{j}$  & $i_v > \Theta_{j}$ \\
        \hline
$m_\text{rem} \geq M_\text{min} $ &  1.  & 0. \\
$m_\text{rem} < M_\text{min} $ & 0. & 1. \\
    \end{tabular}
    \caption{Our construction for the electromagnetic likelihood ($\mathcal{L}(j|\theta, \Lambda)$) of a hierarchical sample $\Lambda$($M_\text{min}, \theta_\text{max}$) given the inclination angle and remnant mass posteriors of the event parameter estimation.}
    \label{tab:jet_likeihood}
\end{table}

\section{results}
\label{sec:results}
We perform population inference on the population of NSBH mergers using the hierarchical model described in Section~\ref{sec:maths}. We sample uniformly in the threshold remnant mass $M_\text{min}$ between $\unit[0.0001-0.2]{M_\odot}$ and uniformly in the jet opening angle $\Theta_j$. We sample with the nested sampling package Nestle \citep{Barbary2021}. 

\begin{figure}
    \centering
    \includegraphics[width=0.95\columnwidth]{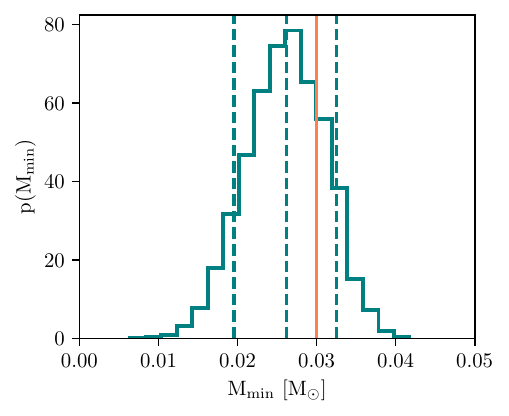}
    \caption{Posterior distribution for the threshold remnant mass for \nNSBH NSBH events detected with \asharp sensitivity. The dashed lines denote the 10, 50 and 90 percent credible intervals. The solid orange line denotes the true value of $\unit[0.03]{M_\odot}$. The true value is recovered accurate to within $\unit[0.01]{M_\odot}$ at the 90\% credible interval.  }
    \label{fig:mthresh_only}
\end{figure}

\begin{figure}
    \centering
    \includegraphics[width=\columnwidth]{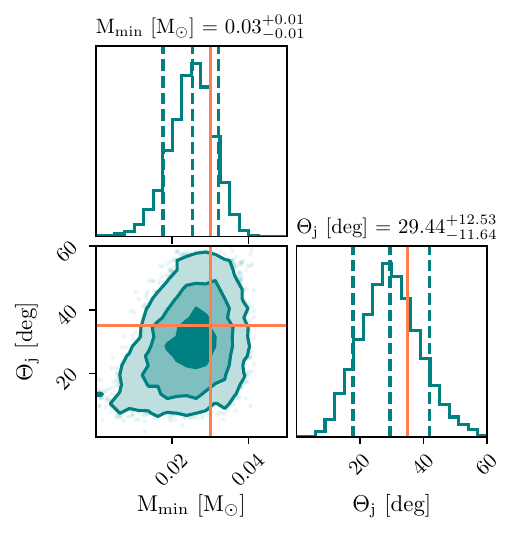}
    \caption{Two dimensional posterior distribution for the threshold remnant mass and threshold viewing angle. The solid orange lines denote the true values of $\unit[0.03]{M_\odot}$ and $\unit[35]{^\circ}$. We recover the threshold disk mass to within $\unit[0.01]{M_\odot}$ at the 90\% credible interval and the threshold viewing angle to within $\unit[13]{^\circ}$.}
    \label{fig:2d_posterior}
\end{figure}

We first sample in $M_\text{min}$ only, assuming a known jet opening angle of $\unit[35]{^\circ}$. 
Figure~\ref{fig:mthresh_only} shows the results of our single parameter inference on $M_\text{min}$. We find that we can recover $M_\text{min}$ to within $\unit[0.01]{M_\odot}$ at the 90\% credible interval level. Next, we relax our assumption of a known jet opening angle and sample in the jet opening angle as well. These results are shown in Figure~\ref{fig:2d_posterior}. We find that $M_\text{min}$ is recovered almost as well when sampling over $\Theta_j$. Additionally, $\Theta_j$ is recovered to within $\unit[13]{^\circ}$ at the 90\% credible interval. 
Our posteriors appear to show a slight positive correlation between $M_\text{min}$ and $\Theta_j$. We interpret this correlation as a contour of ``constant electromagnetically-bright'', since increasing $M_\text{min}$ decreases the number of electromagnetically-bright binaries, while increasing $\Theta_j$ serves to increase this number. 

\begin{figure}
    \centering
    \includegraphics[width=\columnwidth]{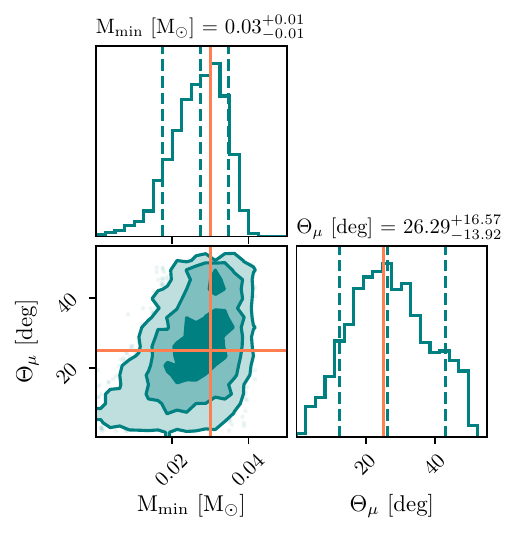}
    \caption{Two dimensional posterior distribution for the threshold remnant mass and mean of the distribution of opening angles. The solid orange lines denote the true values of $\unit[0.03]{M_\odot}$ and $\unit[25]{^\circ}$. We recover the threshold disk mass to within $\unit[0.01]{M_\odot}$ at the 90\% credible interval and the threshold viewing angle to within $\unit[16]{^\circ}$.}
    \label{fig:2d_posterior_mu}
\end{figure}

We relax our assumption that all gamma-ray burst jets will have the same opening angle $\Theta_j$. We draw $\Theta_j$ from a truncated Gaussian distribution between $\unit[2]{^\circ}$ and $\unit[50]{^\circ}$ with mean $\unit[25]{^\circ}$ and width $\unit[5]{^\circ}$. We fix the minimum, maximum and standard deviation and sample in the mean of this distribution $\Theta_\mu$, along with $M_\text{min}$. The results of this analysis are shown in Figure~\ref{fig:2d_posterior_mu}. We obtain a similar precision in our $M_\text{min}$ posterior to the one we obtain with a constant distribution of $\Theta_j$, however our recovery of $\Theta_\mu$ is broadened compared to when we sample in $\Theta_j$. 
More events would be required to regain the precision in jet geometry recovery we obtain in the more simplistic analysis.

 Our results imply that with one year of NSBH detections in \asharp, the GRB jet parameters begin to be well-resolved, even though only five of our simulated NSBH observations had multi-messenger detections. 
This suggests that---given favorable conditions---our method is a viable way to infer population-level jet physics parameters like $M_\text{min}$ as early as the \asharp era and certainly by the time of next generation detectors such as Cosmic Explorer.

\section{Discussion and Conclusions}
\label{sec:discuss}
The physics of NSBH mergers, whether these systems launch jets and the physics of these jets are unknown. We expect that even if multi-messenger NSBH mergers are observed, the jet physics will remain highly uncertain. However, in the coming decade, gravitational-wave astronomy may provide insights into NSBH jets. 
In this \textit{Paper} we study the ability of gravitational-wave observatories to measure the minimum remnant mass needed to launch a jet; $M_\text{min}$ and the jet opening angle implied by a GRB being detected on Earth; $\Theta_j$. Using optimistic yet plausible assumptions about the NSBH population and neutron star equation of state, and minimal dependence on GRB jet modeling, we find that these parameters can be resolved in an \asharp detector configuration with approximately 200 NSBH observations when we assume all GRB jets have a same opening angle (as shown in Figure~\ref{fig:2d_posterior}).
Our measurement of the minimum remnant mass is robust to assumptions about the jet geometry.
Our inferences on the jet opening angle distribution become less precise, but remain informative when we instead measure the mean of a distribution of jet opening angles along with the remnant mass (shown in Figure~\ref{fig:2d_posterior_mu}). 
We note that Cosmic Explorer will observe $\sim 10^4$ NSBH mergers per year out to redshift 10 \citep[e.g.,][]{Gupta2023}. Assuming that next-generation GRB instruments can capture bursts from this distance, and scaling up the electromagnetic detection rate we use in this study, our population of NSBH detections could result in $\sim 250$ electromagnetically bright NSBH mergers per year in the Cosmic Explorer. 
Even accounting for more complicated models of jet geometry, e.g., structured jets, distributions in jet opening angles etc., the volume of data available from Cosmic Explorer will allow us to precisely measure the required disk mass to launch a GRB jet as well as e.g., the parameters of the truncated Gaussian opening angle distribution or more complicated models of jet geometry.

Our analysis makes a number of simplifying assumptions.
We assume a top-hat jet structure characterized by an abrupt cutoff between detection and non-detection that depends on the jet opening angle. 
However, jet geometries are likely to be structured, rather than following a simple top hat model \citep[e.g.,][]{Urrutia2021, Salafia2022}. 
More realistic jet models could be considered in our analysis, such as a structured jet \citep[e.g.,][]{Kathirgamaraju2018}. 
Including systematic uncertainty in the jet profile would cause our posteriors on the parameters we consider in this analysis to get broader, and necessitate more parameters be included in our jet model \citep[e.g.,][]{Howell2019, Sarin2022, Biscoveanu2020}. 
In this sense, our results are optimistic.
We have made another simplifying assumption, treating the top-hat jet luminosity as a step-function in the detector frame. In reality a top-hat jet would display some structure in the jet energy received in the detector frame, as can be seen in e.g., \citet{Biscoveanu2020}. 
In effect, we assume that the jets are launched in a highly relativistic regime, with a Lorentz factor $\gtrsim300$. Although the true relativistic nature of GRB jets remains unknown, there is some evidence that the bulk Lorentz factor could approach 1000 \citep{Abdo2009, Ackermann2010}. We also note that some proposed emission mechanisms of GRB jets, such as cocoon shock breakout \citep[e.g.,][]{Gottlieb2018}, may only be mildly relativistic with Lorentz factors $< 300$.
If we were to marginalize over some model for Lorentz factor, this would broaden our posteriors. 
We have assumed that jets propagate along the axis of total angular momentum. Our analysis would incur additional systematic errors if jets propagate with some misalignment relative to the system total angular momentum \citep[e.g.,][]{Muller2024}.

We have assumed that the neutron star equation of state and the NSBH population distributions are known. These parameters should in principle be marginalized over to account for their uncertainty. We expect that marginalizing over the equation of state uncertainty when calculating $m_\text{rem}$ will broaden our posteriors in $M_\text{min}$. So too will marginalizing over the population parameters, although to a lesser extent, since most of our inference comes from the few events close to $M_\text{min}$ and $\Theta_j \simeq \theta_\text{JN}$, whereas full population inference includes information from all events. We also note that the fitting formula we employ for $m_\text{rem}$ \citep{Foucart2018} quotes an average relative error of 15\% on the baryonic remnant mass, which we expect to be a small contribution compared to the other systematic errors we discuss here. Future analyses should consider including some of these uncertainties. 

Our results imply $\approx \nNSBH$NSBH mergers \textit{co-observed} with gravitational wave and gamma-ray burst instruments are required to constrain $M_\text{min}$ to within $0.01 M_\odot$. We consider that the duty cycle and sky coverage of gravitational-wave and gamma-ray observatories may mean that some NSBH mergers may be mis-classified as non-jet detections. This will effectively increase the observing time needed to constrain the jet parameters to the precision we present here. LIGO \asharp is slated to begin observing in the early 2030s. By that time, several next-generation instruments capable of observing gamma-ray bursts may be observing \citep[e.g.,][]{Bozzo2024} such as CTAO \citep{CherenkovTelescopeArrayConsortium2019}, which is expected to achieve close to all-sky field of view, as well as THESEUS \citep{Amati2018} and HERMES \citep{Fuschino2019}. 
The CTAO is predicted to detect gamma-ray bursts from $\approx 10\%$ of the binary neutron stars detected by LIGO to an inclination of $\unit[30]{^\circ}$ \citep{Mondal2024}. Dedicated radio follow up of GRB afterglows will decrease the fraction of these missed jets \citep{Colombo2024}. At least nine GRB instruments are planned to be operating alongside LIGO \asharp (see, e.g., Table~2 of \cite{Burns2019}), with a range of energies, sky coverages and cadences. 
We optimistically anticipate that most gamma-ray bright NSBH mergers will be identified through at least one instrument in the \asharp era. 

Coincident unrelated transients may contaminate our population of NSBH events, such as through unrelated GRBs arriving in temporal and/or spatial coincidence with a non-disrupting NSBH merger or mis-classification of other gravitational-wave sources as NSBH mergers. 
We expect that the majority of transients will be well-localized in \asharp, reducing the chance of false positive associations.
However one should take care to ensure that multi-messenger detections are probabilistically favored using e.g., the Bayesian odds statistic \citep[e.g.,][]{Ashton2018, Clarke2024}. 

Gravitational-wave signals originating from binary neutron stars, binary black holes, or even non-astrophysical noise artifacts could be misidentified as NSBH mergers and contaminate our population. However, we expect the amount of contamination to be small. Black holes in the lower-mass gap misidentified as neutron stars may be the largest source of contamination, particularly if there is considerable overlap in the mass distributions of black holes and neutron stars \citep[e.g.,][]{Littenberg2015}.
The contaminants could serve to increase the systematic uncertainty on our inference of $M_\text{min}$, but we expect the impact to be small compared to other uncertainties. Improvements to individual source modeling and binary population modeling will decrease the chances of contaminating our NSBH population \citep[e.g.,][]{Chen2020, Coupechoux2022, Golomb2024}. 

By the 2030s, when we expect to have enough NSBH mergers to perform this analysis, our knowledge about the neutron star equation of state and the profile of gamma-ray burst jets may be sufficient to minimize systematic error in this analysis. 
Even taking into account systematic error, we are optimistic that the constraints presented in this \textit{Paper} are sufficiently narrow that near-future constraints on NSBH jets from \asharp will be interesting. The prospects for our analysis in the Cosmic Explorer era are even more promising.

\section{Acknowledgments}

We thank the referee for their helpful suggestions. 
We thank Shanika Galaudage and Nikhil Sarin for their helpful comments on this manuscript. 
This work is supported through Australian Research Council (ARC) Centres of Excellence CE170100004 and CE230900016, Discovery Projects DP220101610 and DP230103088, and LIEF Project LE210100002. 
T. A. C. receives support from the Australian Government Research Training Program.
This material is based upon work supported by NSF's LIGO Laboratory which is a major facility fully funded by the National Science Foundation.
The authors are grateful for for computational resources provided by the LIGO Laboratory computing cluster at California Institute of Technology supported by National Science Foundation Grants PHY-0757058 and PHY-0823459, and the Ngarrgu Tindebeek / OzSTAR Australian national facility at Swinburne University of Technology.

\section{data availability}

The data underlying this article will be shared on reasonable request to the corresponding author.

This work made use of the following software packages: \texttt{astropy} \citep{astropy:2013, astropy:2018, astropy:2022}, \texttt{matplotlib} \citep{Hunter:2007}, \texttt{numpy} \citep{numpy}, \texttt{python} \citep{python}, \texttt{Numba} \citep{numba:2015, Numba_11642058}, \texttt{pandas} \citep{mckinney-proc-scipy-2010, pandas_13819579}, \texttt{scipy} \citep{2020SciPy-NMeth, scipy_11702230}, \texttt{Bilby} \citep{Ashton_2019_bilby, Bilby_2602178}, \texttt{dynesty} \citep{dynesty}, \texttt{nestle} \citep{Barbary2021} and \texttt{corner.py} \citep{corner-Foreman-Mackey-2016, corner.py_4592454}.
Software citation information aggregated using \texttt{\href{https://www.tomwagg.com/software-citation-station/}{The Software Citation Station}} \citep{software-citation-station-paper, software-citation-station-zenodo}.

\bibliographystyle{aasjournal} 
\bibliography{bib}
\end{document}